\documentclass{conm-p-l}
\usepackage[dvips]{changebar}
\usepackage[dvips]{color}
\usepackage{amssymb,amsthm}

\usepackage{a4}
\numberwithin{equation}{section}

\usepackage[mathscr]{eucal}

\usepackage{version}
\excludeversion{LONG}

\newcounter{mnotecount}[section]
\renewcommand{\themnotecount}{\thesection.\arabic{mnotecount}}

\newcounter{mymnotecount}[section]
\renewcommand{\themymnotecount}{\thesection.\arabic{mymnotecount}}
\newcommand{\mymnote}[1]
{\protect{\stepcounter{mymnotecount}}$^{\mbox{\footnotesize $
\bullet$\themnotecount}}$ \marginpar{
\raggedright\tiny\em $\!\!\!\!\!\!\,\bullet$\themymnotecount: #1} }
\renewcommand{\mymnote}[1]{}

\def\ben{\begin{equation}}
\def\een{\end{equation}}
\newcommand{\bea}{\begin{eqnarray}}
\newcommand{\eea}{\end{eqnarray}}

\theoremstyle{plain}
\newtheorem{thm}{Theorem}[section]

\begin{document}
\date{\today \ {\em File:\jobname{.tex}\,}}

\title{The stationary n-body problem in general relativity}

\author[R. Beig]{Robert Beig${}^{\dagger}$} \email{robert.beig@univie.ac.at}
\address{Vienna University, Gravitational Physics, Faculty of Physics, Boltzmanngasse 5, A-1090 Vienna, Austria}
\thanks{${}^{\dagger}$ Supported in part by Fonds zur F\"orderung der
Wissenschaftlichen Forschung project no. P20414-N16.}

\setcounter{tocdepth}{2}

\begin{abstract}
In this talk I describe recent joint works with R.Schoen and
with G.Gibbons and R.Schoen which prove the non-existence
of certain asymptotically flat, stationary solutions of the
Einstein equations with more than one body. The basic restriction
is for example satisfied when spacetime has an isometry reversing the
sign of the timelike Killing vector and fixing a hypersurface
in the space of Killing trajectories which is disjoint from the bodies.
I also give a detailed treatment of the Newtonian situation.

\end{abstract}

\maketitle
\numberwithin{equation}{section}
\renewcommand{\theequation}{\thesection.\arabic{equation}}
\section{Introduction}

'A configuration of several gravitating bodies can not be in stationary equilibrium':
this plausible expectation is in the context of GR still awaiting a rigorous proof. We will in this talk start by explaining
the Newtonian situation. This can be summarized by the statement that there
are no static solutions with more than one body, and where there exists a plane separating these bodies.
This statement is well-known, but I have not seen a printed version. That some separation condition is needed in the statement
of that result is clear from the fact that without it there are in fact solutions for the static 2 body problem, with source
an elastic solid, both in the Newtonian \cite{bs} and in the Einstein theory \cite{as}, for example with one small body put near
a critical point of the potential of a large body, this critical point lying in a suitably hollow region near the large body.\\
In the next section we consider a nonlinear generalization of the Newtonian theory due to Giulini. Here we are able to obtain
a weaker result in the spirit of the general relativistic result in \cite{bgs}. Namely, we show there can not exist solutions
where the potential has zero normal derivative on a plane lying in the vacuum region. This result, which is in the spirit of
 the GR result in \cite{bgs}, in particular rules out
several-body solutions which are reflection symmetric across a plane in the vacuum region. In the final section we describe
one of the results in \cite{bgs}. This states that there can be no stationary spacetime having a reflection symmetry which
inverts the orientation of the orbits of the timelike Killing vector and maps a timelike hypersurface tangential to these orbits into itself. This corresponds to a situation where the spin-spin interaction between matter on the two sides
of this timelike hypersurface is attractive. \\
In the case of two axially symmetric black coaxial black holes the problem treated here has also been studied \cite{w94},
and an analogous result has been proved in \cite{lt}. In the absence of the discrete symmetry we are imposing, i.e. where
there is the a priori possibility for spin-spin forces to balance the gravitational attraction, a full proof of nonexistence has
been found very recently in \cite{nh}, using methods from the theory of completely integrable systems.
\section{The Newtonian situation}
Let $\Omega_{(\alpha)}$ ($\alpha=1,2,..N$) be bounded open connected sets with smooth boundary $\partial \Omega_{(\alpha)}$
in $\mathbb{R}^3$ with Euclidean metric $\delta_{ij}$. The sets $\Omega_{(\alpha)}$ play the role of support of bodies.
Each body has associated with it a positive function $\rho_{(\alpha)}$, the mass density, and the Cauchy stress tensor, a symmetric tensor field
$\sigma_{\!(\alpha)ij}$. In a theory of continuum mechanics such as ideal elasticity, the quantities $\rho_{(\alpha)}, \sigma_{\!\!(\alpha)ij}$ are given, not as functions on $\mathbb{R}^3$, but as functions of some further fields such as maps
from $\mathbb{R}^3$ into some 'material manifold' $\mathcal{B}_\alpha$ and of the first derivatives of these maps.
The basic equations then take the following form ($\Delta = \delta^{ij}\partial_i \partial_j$)
\begin{equation}\label{U}
\Delta U = 4 \pi G \sum_\alpha \rho_{(\alpha)}\, \chi_{\Omega_{(\alpha)}}\,,\hspace{0.7cm}U \rightarrow 0\,\, \mathrm{at}\,\,\infty
\end{equation}
with $\rho_{(\alpha)} > 0$ and
\begin{equation}\label{sigma}
\partial^j \sigma_{\!(\alpha)ij} = \rho_{(\alpha)}\partial_i U\hspace{0.3cm}\mathrm{in}\,\,\Omega_{(\alpha)}\,,\hspace{0.7cm}
\sigma_{\!(\alpha)i}^j n_j|_{\partial \Omega_{(\alpha)}} = 0
\end{equation}
The detailed structure of the matter equations (\ref{sigma}) - which form a quasilinear second-order system of equations, usually elliptic, for the underlying map with Neumann-type boundary condition - need not occupy us here. The important thing is that these equations
imply
\begin{equation}\label{total}
M_\alpha(\xi,U) \Doteq \int_{\Omega_{(\alpha)}}\!\!\!\!\! \xi^i \partial_i U \,\,dx = 0\,,
\end{equation}
where $\xi^i$ is any of the 6 Euclidean Killing vector, i.e. satisfies $\partial_{(i}\,\xi_{j)}=0$. The meaning of (\ref{total})
is of course that the total force and the total torque acting on each body be zero. One easily deduces (\ref{total})
from (\ref{sigma})
by using integration by parts, the symmetry of the Cauchy stress together with the Killing equation and the boundary condition
of vanishing normal stress in (\ref{sigma}).\\
The next question is if perhaps Eq.(\ref{U}) alone implies these conditions. The answer is 'yes' for $N=1$.
Namely the 'Newtonian stress tensor' defined by
\begin{equation}\label{theta}
\Theta_{ij} = \frac{1}{4 \pi G}\,[(\partial_i U)(\partial_j U) - \frac{1}{2}\delta_{ij}(\partial U)^2]
\end{equation}
satisfies, by virtue of (\ref{U}), the relation
\begin{equation}\label{distr}
\partial^j \Theta_{ij} = \sum_\alpha \rho_{(\alpha)} \chi_{\Omega_{(\alpha)}}\, \partial_i U\,.
\end{equation}
Thus the gravitational equation (\ref{U}) together with the matter equations (\ref{sigma}) -
including the boundary conditions - imply that
\begin{equation}\label{distr1}
\partial^j(\Theta_{ij} - \sum_\alpha \sigma_{\!(\alpha)ij} \chi_{\Omega_{(\alpha)}})= 0
\end{equation}
holds in the sense of distributions.\\
Integrating (\ref{distr}) against $\xi^i$ over $\mathbb{R}^3$ we get zero on the l.h. side: integrate by parts and use
the Killing equation. The boundary term gives zero, since $\partial U = O(|x|^{-2})$. Thus
\begin{equation}\label{sum}
\sum_\alpha M_\alpha(\xi,U) = 0
\end{equation}
In fact, defining $U_{(\alpha)}$ to be the unique solution of
\begin{equation}\label{Ualpha}
\Delta U_{(\alpha)} = 4 \pi G \rho_{(\alpha)}\, \chi_{\Omega_{(\alpha)}}\,,\hspace{0.7cm}U_{(\alpha)} \rightarrow 0\,\, \mathrm{at}\,\,\infty
\end{equation}
we find, by applying the same argument first to $U_{(\alpha)}$ and then to $U_{(\alpha)} + U_{(\beta)}$, that
\begin{equation}\label{actreact}
M_\alpha(\xi,U_{(\beta)}) + M_\beta(\xi,U_{(\alpha)}) = 0
\end{equation}
This says that the total force and torque exerted by the gravitational field of body 1 on body 2 is minus that
exerted by body 2 on body 1 - which of course is nothing but the law 'actio = reactio'. Eq.(\ref{actreact})
also entails, for $\alpha = \beta$, the statement that the self-force and self-torque due to the gravitational field of a body vanishes.\\
We now observe that (\ref{actreact}) still falls short of guaranteeing (\ref{total}) when $N>1$, because, for $N=2$ say,
we need, in order for (\ref{total}) to hold, in addition to $M_1(\xi,U_{(1)})$, $M_2(\xi,U_{(2)})$ and $M_1(\xi,U_{(2)}) + M_2(\xi,U_{(1)})$ all being zero,
that also $M_1(\xi,U_{(2)})$ and $M_2(\xi,U_{(1)})$ vanish separately.\\
Now, finally, we show that $M_1(\xi,U_{(2)})$ is indeed non-zero when $\Omega_{(1)}$  and $\Omega_{(2)}$ are separated by a plane.
More precisely, when $\Omega_{(1)}$ lies to the left and $\Omega_{(2)}$ to the right of some separating plane $S$,
we will show that
\begin{equation}\label{sign}
\partial_n U_{(2)}|_\Sigma > 0\,
\end{equation}
on every plane $\Sigma$ parallel to and to the left of $S$ and where $n$ is the left normal. This, by (\ref{total}), will clearly imply $M_1(\xi,U_{(2)})>0$,
 when $\xi$ is taken to be a translational Killing vector pointing along the left-normal of $\Sigma$. Inequality (\ref{sign}) is nothing but the statement that the gravitational force
due to body 2 on any body separated from body 2 by a plane is attractive. To prove (\ref{sign}), we use an elegant argument
due to R.Schoen. Consider, in region 2, i.e. to the right of $\Sigma$, the field $\bar{U}(x) = U_{(2)}(x) - U_{(2)}(\rho \circ x)$, where
$\rho$ is the reflection across $\Sigma$. The field $\bar{U}$ satisfies
\begin{equation}\label{region2}
\Delta \bar{U} = 4 \pi G \rho_{(\alpha)}\chi_{\Omega_{(\alpha)}}\hspace{0.4cm}\mathrm{in\,\,region\,\, 2}
\end{equation}
Furthermore $\bar{U}$ goes to zero at infinity and vanishes on $\Sigma$. Thus, by the maximum principle, it follows that
$\bar{U} < 0$. Then, by the Hopf lemma, $\partial_n \bar{U}|_\Sigma = 2 \,\partial_n U_{(2)}|_\Sigma > 0$. We have thus proved the\\
\begin{thm} There exists no solution to the equations (\ref{U},\ref{sigma}) with $N>1$ and where there is
a plane $S$ separating the bodies.
\end{thm}
Let us note that all the results obtained here could have been proved using the explicit Green's function of the flat Laplacian,
namely that, for compactly supported $\rho$, the unique solution $U$ of $\Delta U = 4 \pi G \rho$ with $U$ decaying
at infinity has the form
\begin{equation}\label{explicit}
U(x) = - \,G \int_{\mathbb{R}^3} \frac{\rho(x')}{|x - x'|}\,\, dx'
\end{equation}
We have avoided this and tried to make the argument as conceptual as possible. It is clear for example that the above result
remains to be true, when the Laplacian $\Delta$ is replaced by 'meson-type' operator $\Delta - \mu^2$, with $\mu = \mathrm{const}$.
We have however made crucial use of the linearity of the Poisson equation, which allows to consider the concept of 'force
on a body due to the gravitational field of some other body'. This is not any longer available in the following
nonlinear modification of the Newtonian theory.
\section{The Giulini theory}
This is a nonlinear modification of the Newtonian theory where $1/c^2$ of the gravitational self-energy  has an active gravitational mass associated with it (the constant $c$ being the speed of light). The equations now are
\begin{equation}\label{g1}
\Delta U = \frac{4 \pi G}{c^2}\,\sum_\alpha \epsilon_{(\alpha)}\chi_{\Omega_{(\alpha)}} - \frac{1}{2c^2}\,(\partial U)^2\,,\hspace{0.7cm}U \rightarrow 0\,\, \mathrm{at}\,\,\infty
\end{equation}
\begin{equation}\label{g2}
e^{-\frac{U}{c^2}}
\partial^j \left(e^{\frac{U}{c^2}}
\sigma_{(\alpha)ij}\right) = \frac{1}{c^2}\,\, \epsilon_{(\alpha)}\,\partial_i U\hspace{0.3cm}\mathrm{in}\,\,\Omega_{(\alpha)}\,,\hspace{0.7cm}
\sigma_{\!(\alpha)i}^j n_j|_{\partial \Omega_{(\alpha)}} = 0
\end{equation}
with $\epsilon_{(\alpha)} > 0$ playing the role of 'rest energy plus internal energy density'. The
equation (\ref{g1}) can be written in the linear form $\Delta \Psi = \frac{2 \pi G}{c^4}\, \Psi\,\sum_\alpha \epsilon_{(\alpha)}\chi_{\Omega_{(\alpha)}}$ in terms of $\Psi = e^\frac{U}{2 c^2}$, but that will play no role for us.
The potential $U$ has the expansion
\begin{equation}\label{U1}
U = - \frac{G m}{r} + O^\infty(\frac{1}{r^2})
\end{equation}
The constant $m$ in (\ref{U1}) is positive. To prove this, observe that
\begin{equation}\label{sphere}
m = \frac{1}{4 \pi G}\lim_{R \rightarrow \infty} \int_{S_R}\!\! e^\frac{U}{2 c^2}\,\partial_i U\,dS^i\,
\end{equation}
where $S_R$ is a large sphere. Now use the Gauss theorem together with (\ref{g1}) on the r.h. side of (\ref{sphere}).\\
Again, in this theory, there
is a law 'actio = reactio' in the form
\begin{equation}\label{g3}
\partial^j\, [e^\frac{U}{c^2}(\Theta_{ij} - \sum_\alpha \sigma_{\!(\alpha)ij} \chi_{\Omega_{(\alpha)}})]= 0
\end{equation}
Now let $S$ be a plane disjoint from all bodies. We can then integrate (\ref{g3}) against $\xi^i$
over one of the half-spaces bounded by $S$, where $\xi$ is the translational Killing vector
coinciding with the unit normal $n$ of $S$. Using integration by parts together with (\ref{U1}) and (\ref{theta}) there results
\begin{equation}\label{g4}
\frac{1}{2} \int_S e^\frac{U}{c^2}\,[(\partial_n U)^2 - (\partial^A U)(\partial_A U)]\, dS = 0\,
\end{equation}
where $\partial_A$ is a partial derivative tangential to $S$. Suppose that $\partial_n U|_S = 0$. It then follows that
$S$ is an equipotential for $U$. But this is impossible due to (\ref{U1}) and the fact that $m \neq 0$. Thus we have obtained the\\
\begin{thm} If a solution of (\ref{g1},\ref{g2}) contains a plane $S$ disjoint from the bodies
on which $\partial_n U$ is zero, this has to be the trivial ('no-body'-) solution with $U=0$.
\end{thm}
The hypothesis of the above theorem is fulfilled when there is a reflection across some plane $S$ disjoint from the bodies
which maps the set of bodies into itself. It is this theorem - which is weaker than that available in the 'pure' Newtonian case - for which
there exists an analogue in GR, to which we now turn.
\section{The GR situation}
The result I am now explaining works for the stationary Einstein equations with sources compactly supported in space.
The nature of these sources is completely irrelevant except that they should meet the requirements of
the positive-energy theorem, i.e. obey the dominant-energy condition\footnote{Actually, all we need is that the ADM energy in the
asymptotic rest system of the Killing vector
 be nonzero.}. Otherwise the field equations
will only be used in the vacuum region. For simplicity we will restrict to $3+1$ dimensions, although
an analogous results for $n+1$ with $n>3$ is also true. We consider stationary - not just static - case, to allow
for spin-spin interactions. Thus we assume $(M, ds ^2)$ has a timelike Killing vector $\xi^\mu$ with
complete orbits. Then $ds^2$ can be written as (in distinction to the previous section, we set $c$ equal to 1)
\begin{equation}\label{metric}
ds^2  = - e^{2 U} (dt + \psi_i dx^i)^2 + e^{- 2 U}\,
h_{ij} dx^i dx^j
\end{equation}
with $h$ being a Riemannian metric on $N$, the quotient space under the action of $\xi = \partial_t$.
We refer to $U$ as the gravitational potential and define $1/2$ the curvature of the Sagnac connection $\psi$ by
\begin{equation}\label{sagnac}
\omega = \frac{1}{2}\, d \psi,\hspace{1.2cm}\omega_{ij} = \partial_{[i} \psi_{j]}
\end{equation}
the vacuum field equations turn out to be ($\mathcal{G}_{ij} =
\mathcal{R}_{ij} - \frac{1}{2} h_{ij} \mathcal{R}$)
\begin{subequations}\label{field}
\begin{align}
\mathcal{G}_{ij} - 8 \pi G\,(\Theta_{ij} + \Omega_{ij})&= 0\\
\Delta_h U + e^{4 U}\,\omega_{kl}\omega^{kl}&=0\\
D^j\left(e^{4 U} \,\omega_{ij}\right)& = 0
\end{align}
\end{subequations}
In (\ref{field}a) we are using the definitions
\begin{equation}\label{def1}
8 \pi G \,\Theta_{ij} = 2\, [(D_i U) (D_j U) -
\frac{1}{2}h_{ij}(DU)^2]
\end{equation}
and
\begin{equation}\label{def2}
8 \pi G\,\Omega_{ij} = 2 \,e^{4 U}\,[\,
- \omega_{ik}\omega_{j}{}^k + \frac{1}{4} \,h_{ij}\,
\omega_{kl}\omega^{kl}]
\end{equation}
(Using $d \omega = 0$, Eq.(\ref{field}c) follows from Eq.'s (\ref{field}a,b).) We consider asymptotically flat solutions of (\ref{field}).
They can be shown \cite{bsi} to have the form
\begin{subequations}\label{as}
\begin{align}
h_{ij}&=\delta_{ij} + O^\infty\left(\frac{1}{r^2}\right)\\
U &= - \frac{G m}{r} + O^\infty\left(\frac{1}{r^2}\right)\\
\omega_{ij} &= - G\, \frac{L_{ij} + 3
\,a_{[i}L_{j]k}a^k}{r^3} + O^\infty\left(\frac{1}{r^4}\right)
\end{align}
\end{subequations}
in suitable coordinates. Here $a^i = \frac{x^i}{r}$.
Furthermore the constants $m$ and $L_{ij}=L_{[ij]}$ are respectively
the ADM mass and the (mass-centered) spin tensor of the
configuration. When $m=0$, the ADM energy of the initial data set
induced on $t=\mathrm{const}$ is also zero. Then, when the source
satisfies the dominant energy condition, the positive energy theorem
implies that spacetime is flat\footnote{One can also
allow for the presence of horizons, see \cite{ghp},\cite{bc}.}.
\begin{thm} Let $(N,h)$ have a hypersurface $S$ disjoint from
the matter region, which is non-compact, closed and totally geodesic
w.r. to the unrescaled metric $e^{- 2 U} h_{ij}$.
Suppose in addition that the pull-back-to-$S$ of
$\omega$ is zero. Then spacetime is flat.
When spacetime has more than one asymptotically
flat end, all the statements above hold separately
w.r. to any such end.
\end{thm}
We first have to comment on the surface $S$ in the hypothesis of the theorem.
The typical situation where our hypothesis on $S$ will be satisfied is
when there is a reflection isometry $\Psi$ of spacetime which maps $\xi$ into $ - \xi$ (and thus
projects down to an isometry of $(N, h_{ij})$ and $(N, e^{- 2 U} h_{ij})$) and leaves $S$ invariant
\footnote{In fact the presence of the above isometry also implies $D_n U$ to vanish on $S$ - a property
we do not require in the theorem.}.\\
In \cite{bs} it is shown in detail that $S$ has to have a finite number of ends in the asymptotically flat region and
$S$ approaches a plane there in a precise sense. Thus, for each end of $S$ asymptotic coordinates $(x^1, x^A)$ in (\ref{as}) can be chosen
so that each end of $S$ is given by $x^1 = 0$. It then follows from the hypothesis on $\omega$ that $L_{AB}$ in
(\ref{as}c) vanishes. There are then good physical reasons detailed in  \cite{bgs} to believe
that this corresponds to a situation where the spin-spin interaction between the bodies
on the two sides of $S$ is attractive.\\
For the proof of the theorem we contract (\ref{field}a) with $n^i n^j$, with $n^i$
being the unit normal to $S$ in $(N,h)$. The result is
\begin{equation}\label{nn}
\mathcal{G}_{nn} = 2 \left[\frac{1}{2}\,(D_n U)^2 -
\frac{1}{2}\,(D_A U)(D^A U)\right] - \,e^{4 U} \,\omega_{A n} \omega^A{}_n\,\,.
\end{equation}
Here $D_A$ is the intrinsic derivative on $S$. In (\ref{nn}) we
have used that $\omega_{AB}$ is zero. Now, from the Gauss
equation,
\begin{equation}\label{gauss}
\mathscr{R} = - 2\, \mathcal{G}_{nn}  + (\mathrm{tr}\, k)^2 -
\mathrm{tr}(k^2)\,,
\end{equation}
where $\mathscr{R}$ is the Ricci scalar of $S$ and $k$ its extrinsic
curvature. Since $S$ is totally geodesic w.r. to $e^{- 2 U} h$, we have that
\begin{equation}\label{bar}
k_{AB} = \,\, q_{AB} \,D_n U
\end{equation}
where $q$ is the intrinsic metric on $S$ induced from $h$. Inserting
(\ref{bar},\ref{nn}) into Eq.(\ref{gauss}), the terms involving $D_n
U$ cancel so that finally
\begin{equation}\label{pos} \mathscr{R} = 2\, (D_A
U)(D^A U) + 2 \,e^{4 U}\,\omega_{A n}\omega^A {}_n\;.
\end{equation}
In particular $\mathscr{R}$ is non-negative. One now integrates (\ref{pos}) over $S$. Using the version of the Gauss-Bonnet formula
for surfaces with boundary and the asymptotically planar nature of $S$, one arrives at a contradiction except
when $\mathscr{R}$ vanishes and $D_A U$, whence $U$, is zero on $S$. Thus by (\ref{as}) $m$ is zero, whence spacetime is flat,
and we are done. \\
We remark that for the $n+1$-dimensional version of this theorem one uses, instead of the Gauss-Bonnet formula, the positive energy theorem in dimension $n-1$\cite{s}.

\end{document}